\documentclass[preprint]{aastex62}
\usepackage{amsmath}
\usepackage{enumitem}[0]

\usepackage[utf8]{inputenc}
\usepackage{tikz}
\usetikzlibrary{shapes.geometric, arrows}
\usepackage{verbatim}
\usepackage[normalem]{ulem}
\usepackage{tabu}

\shorttitle{Exoplanet Exergy}
\shortauthors{Scharf}

\begin{document}

\title {Exoplanet Exergy: why useful work matters for planetary habitabilty}
\author{Caleb Scharf}
\affiliation{Department of Astronomy, Columbia University, 550 West 120th St., New York, NY 10027, USA}
\submitjournal{The Astrophysical Journal 01/03/19, accepted 03/23/19}

\begin{abstract}
The circumstellar habitable zone and its various refinements serves as a useful entry point for discussing the potential for a planet to generate and sustain life. But little attention is paid to the {\em quality} of available energy in the form of stellar photons for phototrophic (e.g. photosynthetic) life. This short paper discusses the application of the concept of exergy to exoplanetary environments and the evaluation of the maximum efficiency of energy use, or maximum work obtainable from electromagnetic radiation. Hotter stars provide temperate planets with higher maximum obtainable work with higher efficiency than cool stars, and cool planets provide higher efficiency of radiation conversion from the same stellar photons than do hot planets. These statements are independent of the details of any photochemical and biochemical mechanisms and could produce systematic differences in planetary habitability, especially at the extremes of maximal or minimal biospheres, or at critical ecological tipping points. Photoautotrophic biospheres on habitable planets around M-dwarf stars may be doubly disadvantaged by lower fluxes of photosynthetically active photons, and lower exergy with lower energy conversion efficiency.
\end{abstract}

\keywords{astrobiology --- planets and satellites: fundamental parameters --- planets and satellites: terestrial planets}

\section{Introduction}
Assessing the potential surface environment of rocky exoplanets typically involves computing the so-called circumstellar habitable zone (CHZ), or liquid water zone (\citet{Dole1964,Hart1979,Kasting1993}). This yields estimates of possible surface temperature distributions, under a variety of assumptions and simplifications about planetary atmosphere and greenhouse effects, climate dynamics, and geophysical cycling (e.g. the carbon-silicate cycle, \citet{Kasting1993}). Refinements include the consideration of atmospheric loss and ionizing radiation, and long-term geodynamo behavior (e.g. \citet{Driscoll2015}) as well as extensions to non-terrestrial environments (e.g. \citet{Seager2013}). The vast majority of CHZ evaluations at this time are made on the basis of planetary orbit, stellar host properties, and planet mass and/or radius constraints from data. All other factors are theoretical projections. The CHZ therefore serves as a rudimentary, first-pass tool for `grading' planets as being more or less likely to harbor life based on terrestrial templates (\citet{Ramirez2018}).

The question of energy availability to any potential life in planetary environments generally receives less attention. Notable work does include the study of the expected surface spectrum of steller radiation from varying mass stars and planetary atmospheric compositions in the context of photosynthetic systems (e.g. \citet{Kiang2007a,Kiang2007b}). On Earth, photosynthetic life does appear to be the largest direct user of stellar photons, with the largest biological energy budget (especially if plant transpiration processes are included, reaching $\sim 5,000$ TW of power use, \citet{jasechko2013}). However, simply adding up available energy (i.e. chemical or electromagnetic) does not tell us about the {\em quality} of that energy. Specifically, the quantity that really matters for life is the availability of {\em useful energy} for doing work.

This brief paper reviews the general application of a basic, but often overlooked, thermodynamic concept: the exergy of a system. In \S2 below the definition of exergy is discussed. In \S3 expressions are provided for exergy in the case of blackbody radiation and the efficiency of conversion of useful energy into work, and these quantities are computed for a range of stellar masses and CHZ planetary surface temperatures, and in \S4 the broad implications and possible future uses of the exergy concept are discussed.

\section{Entropy and Exergy}

In its classical form, entropy is a measure of the energy that is {\em unavailable} to do work in a closed system. The concept of exergy by contrast is a measure of the {\em maximum} amount of work than can be obtained from a system in reference to its environment \citep{Keenan1951,Rant1956}. In principle this makes exergy a straightforward measure of a system, however the literature on exergy is diverse and sometimes rather impenetrable due to different vantage points in different fields. This is particularly true where exergy is applied in ecological or even economical studies. Nonetheless, a classic and useful definition comes from \citet{Szargut1988} who state: ``Exergy is the amount of work obtainable when some matter is brought to a state of thermodynamic equilibrium with the common components of the natural surroundings by means of reversible processes.".

At first glance exergy seems to be equivalent to the Gibbs free energy, which is the amount of available work for an isothermal and isobaric process. But, as above, strictly speaking exergy is a measure of total available work until a system reaches equilibrium with the surroundings, which is not dependent on a process being isothermal or isobaric. So while the Gibbs free energy does not explicitly depend on the system surroundings, exergy does. 

A general form for exergy can be given as:
\begin{equation}
    E_x=U_0+PV_0-TS_0-\sum_{i} \mu_{oi}M_i \;\;,
\end{equation}

where $U_0$ is the internal energy of a system, $P$ is the pressure in the surroundings, $V_0$ is the system volume, $T$ is the temperature in the surroundings, $S_0$ is the system entropy, $\mu_{0i}$ is the chemical potential in the surroundings for a given chemical species in the system, and $M_i$ are the moles present of that species.

Exergy is an explicit application of the 2nd law of thermodynamics, but unlike energy, exergy can be destroyed making it a useful concept for evaluating the efficiency of energy utilizing systems. Evaluating exergy in a system also involves a treatment of the system's reference state environment. This is one reason why exergy sees use in ecological studies and work on human energy use and environmental change (e.g. \citet{chen2005}). Or to put this another way, since exergy is destroyed as work is done it's a useful way to track the real cost of both ecological and human systems.

In the case of radiation fields it is conceptually useful to recognize that a system (e.g. a biological entity) will always be losing energy by radiating due to its finite temperature, while it may also absorb energy from an external field. Exergy is a measure of the balance between this energy loss and gain that accounts for the usability of energy.

\section{Exergy and stellar temperature}

Just as luminosity $L_{\lambda}(T)$ and entropy $S_{\lambda}(T)$ distributions can be computed for Planck's law of a blackbody (\cite{Planck1914, Rosen1954}), so can the exergy distribution $E_{x\lambda}(T)$ for a blackbody system immersed in a blackbody radiation field (see for example \citet{Petela1964,Delgado2017}) since, from Eqn 1., for photons: 

\begin{equation}
E_{x\lambda}=L_{\lambda}(T)-L_{\lambda}(T_0)-T_0(S_{\lambda}(T)-S_{\lambda}(T_0))
\end{equation}

The whole spectrum exergy (i.e. integrated over all wavelengths) for a blackbody at temperature $T_0$ bathed in a blackbody field of temperature $T$ is given by (e.g. \citet{Candau2003}):
\begin{equation}
    E_x^{BB}=\sigma(T^4-\frac{4}{3}T_0T^3+\frac{1}{3}T_0^4)
\end{equation}

To first-order a planet or star can be treated as a blackbody. We can therefore apply this expression to estimate upper limits to both the exergy available from stellar photons at a planetary surface and the total exergy available from stellar photons for the planet. In the former case the relevant blackbody sink temperature is that of the local, surface environment ($T_{s}$) in the latter it is the overall effective temperature of the planet ($T_{eff}$). The difference between these exergy values is also an indication of the exergy used in maintaining the thermal and kinetic state of the planet, i.e. radiative transfer and physical energy transport. For $T_s$ the values used here (273K and 373K) correspond to the `temperate' range for surface environment, where liquid water can exist with a 1 bar pressure, Earth-normal atmospheric composition (the nominal CHZ). For $T_{eff}$ the range depends on the assumed details of atmospheric composition and structure which are expected to differ at the inner and outer edges of the CHZ. Here the range of 175K to 270K is used, based on the results of \citet{Kaltenegger2011}. For more discussion of the CHZ and effective planetary temperatures see \citet{Kopparapu2013} and \citet{Ramirez2018}. Exergy is plotted in Figure 1 as a function of stellar temperature for planets in the CHZ temperature range and for stellar hosts ranging from M8 dwarfs to A5 stars. A solar-temperature star represents a factor $\sim 24\times$ higher exergy source for a CHZ planet than does a low-mass M8 dwarf ($T_*\sim 2660$K).

\begin{figure}
%\epsscale{.80}
\plotone{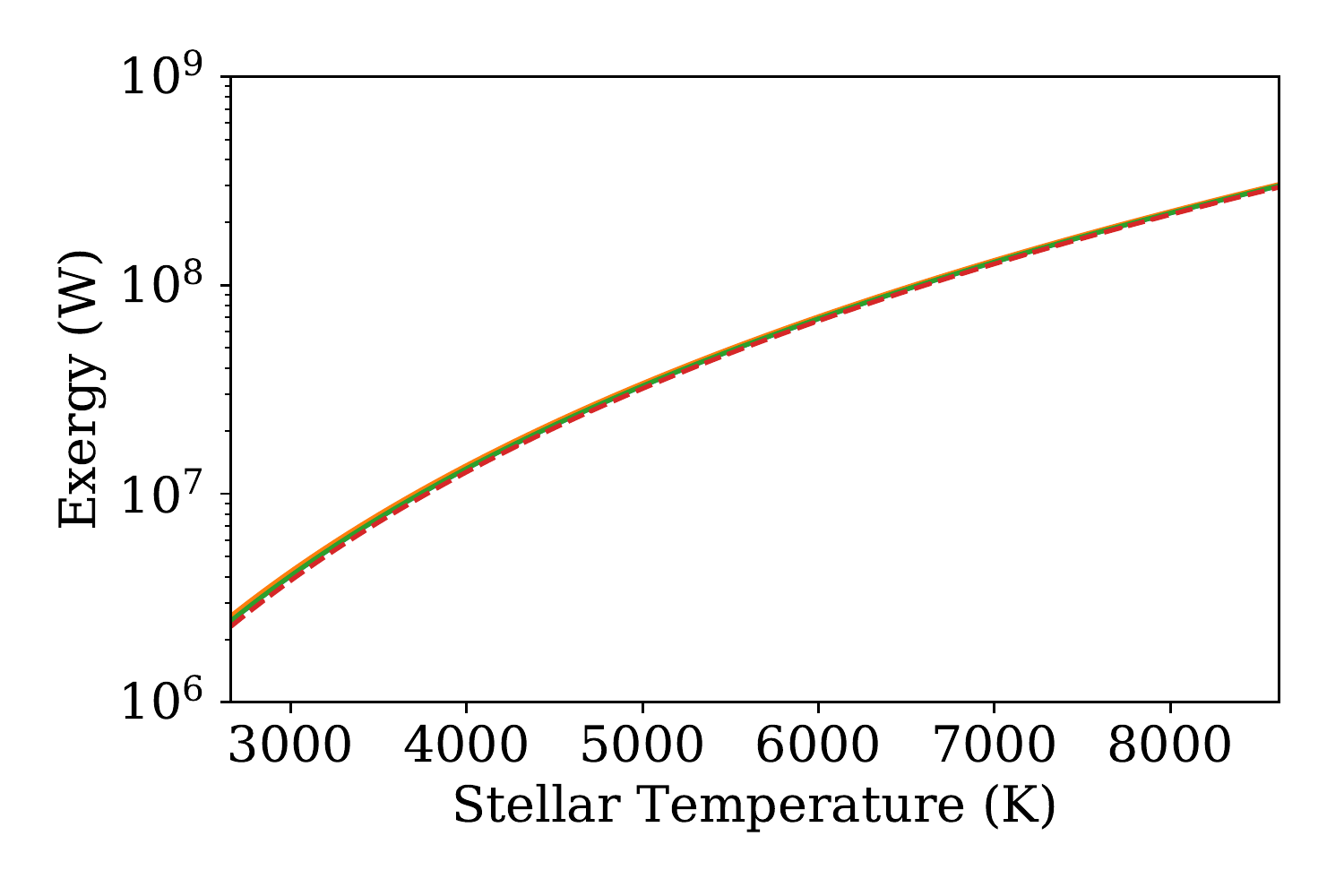}
\caption{Total (whole spectrum) exergy (Watts) with arbitrary normalization as a function of stellar blackbody temperature (from M8 dwarfs at approximately 2660K to A5 stars at approximately 8620K) for surface temperatures $T_{s}=273$K and $373$K and $T_{eff}=175$K and 270K - all curves are effectively indistinguishable on this plot.}
\end{figure}

\subsection{Efficiency}

The 2nd law, whole spectrum efficiency  for the conversion of radiation to work is defined (\cite{Petela1964}) as the ratio of the total (integrated) exergy to the total environmental luminosity (i.e. of the stellar radiation field):
\begin{equation}
    W^{BB}=\frac{E_x^{BB}}{\int_0^{\infty}(L(\lambda,T)d\lambda}=\frac{E_x^{BB}}{\sigma T^4}=1-\frac{4}{3}\frac{T_0}{T}+\frac{1}{3}\left(\frac{T_0}{T}\right)^4
\end{equation}

In Figure 2, $W^{BB}$ is plotted as a function of stellar temperature for planets at the inner ($T_s=373$K) and outer ($T_s=273$K) edge of the CHZ, together with the efficiencies for the corresponding $T_{eff}$ for these planets. The Carnot efficiency, $\eta=1-T_0/T$, is also shown - this represents what is usually the theoretical upper bound on thermodynamic efficiency (and assumes minimal possible entropy increase for work done) irrespective of the form of energy or work.  

\begin{figure}
%\epsscale{.80}
\plotone{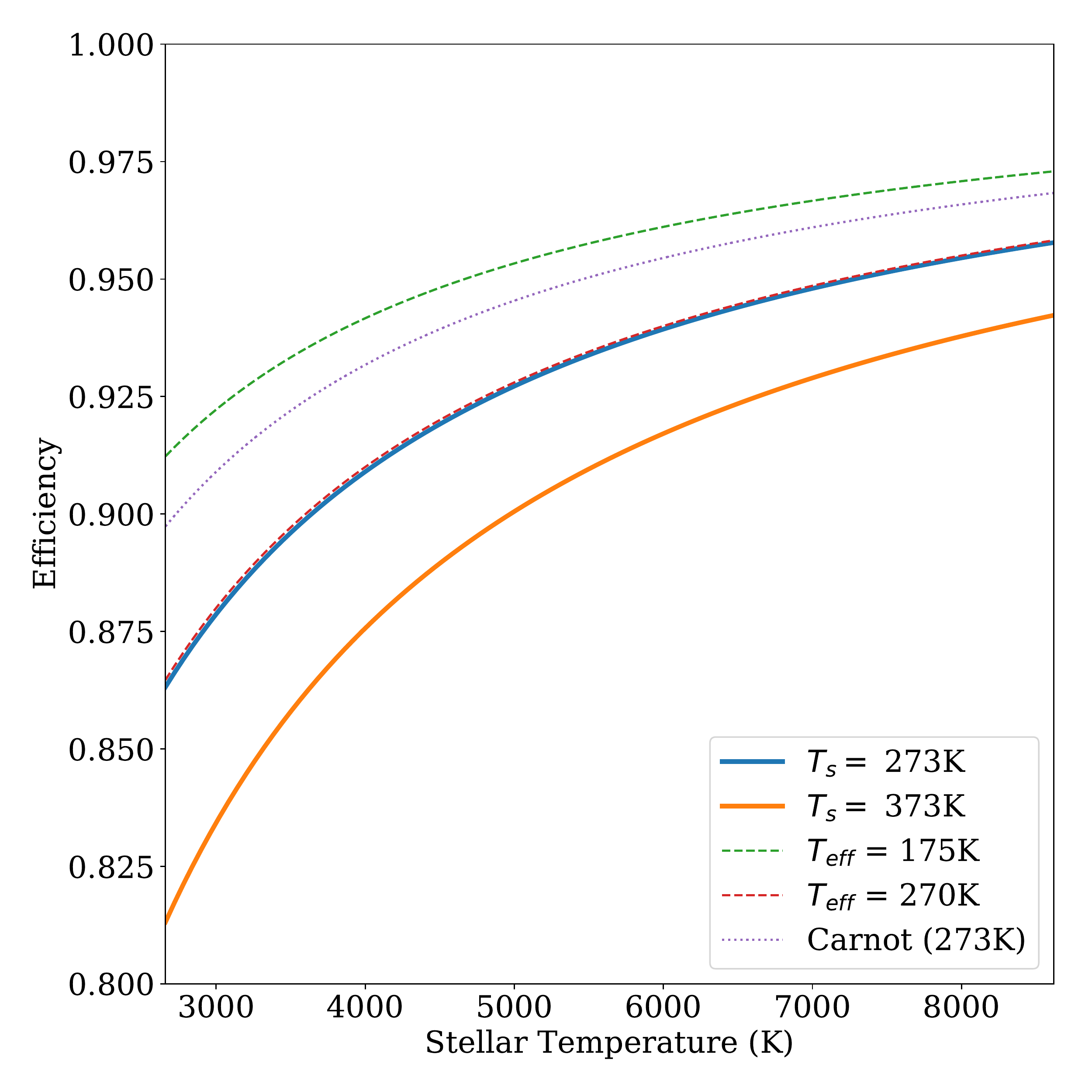}
\caption{Efficiency of the conversion of radiation to work (whole spectrum) as a function of stellar blackbody temperature (from M8 dwarfs at approximately 2660 K to A5 stars at approximately 8620 K) for planetary surfaces with $T_{s}=273$K and $373$K (upper and lower solid curves respectively), and effective temperatures $T_{eff}=175$K and $270$K (upper and lower dashed curves). Carnot efficiency is plotted for $T_0=273$K for comparison (uppermost dotted curve).}
\end{figure}

Surprisingly, there is a noticeable difference in efficiency in conversion to work between the outer and inner temperatures of the classical liquid-water habitable zone, with (for example) $W^{BB}_{outer}/W^{BB}_{inner}\sim 1.06$ for $T_*\sim 2660$K based on the $T_s$ curves. Radiation from a low-mass M-star ($T_*\sim 2660$K) also has a conversion to work efficiency approximately 7\% {\em less} than a Sun-like star ($T_*\sim 5800$K).

It should be noted that these calculations implicitly assume a low cosmic background radiation temperature ($T_{CMB}\sim 3$ K). \citet{loeb2014} has discussed the possibility of liquid water environments on a first generation of planets in the young universe (redshift range $(1+z)=100-137$) where $T_{CMB}=273-373$ K. In that case Equations 3 \& 4 should be modified to include the external cosmic blackbody field in addition to any stellar radiation.

\section{Discussion}

Actual phototrophic systems are unlikely to exploit photons at all wavelengths. For example, on the Earth, the photosynthetically `active' part of the spectrum lies between 400 nm and 700 nm (although there is evidence of certain photosynthetic processes also utilizing near-IR radiation, \citet{nurnberg2018}). The computed efficiency ($W^{BB}$) over a restricted wavelength range is significantly lower. For the above wavelength range on Earth, \citet{Delgado2017} found $W^{BB}\simeq 0.338$ (assuming a planetary temperature of 300K). However, the trend of lower efficiency for cooler stars versus hotter stars remains, irrespective of the assumed spectral band. In simple terms, the closer in temperature two radiation fields are, the less efficiently photons can be converted to work.

For this reason it should also be noted that the {\em local} temperature for any light-harvesting system will impact its theoretical maximum efficiency of conversion of energy to work. For example, thermophilic phototrophic organisms will face an intrinsic, unavoidable shortfall in efficiency (albeit at the level of several percent) compared to organisms in cool environments.  

The exergy values used here assume both a simple blackbody spectrum and no spectral changes due to atmospheric absorption or upper marine environment absorption. In practice the processing of radiation by an atmosphere by irreversible absorption and re-emission must reduce the quality of the radiation reaching the planetary surface, by raising the net entropy above that expected from a pure blackbody field. This will therefore decrease (Eqns. 1 \& 2) the exergy of radiation received at the surface. Thus the same atmosphere that tends to raise and stabilize surface temperatures in the CHZ will also reduce the energy available for useful work at the surface.

An additional caveat is that the calculated efficiency implicitly assumes that all work is carried out while the system is embedded in the external radiation field (e.g. during daytime on the planetary surface). While photosynthesis on Earth immediately `processes' photons into free electrons or protons, that is only the first step in the application of stellar energy for doing useful work. Subsequent chemical steps of varying efficiency will follow, with energy dissipation ultimately sustaining the equilibrium temperature of the biosphere system, $T_0$.

For different stellar hosts, how will these relatively modest ($\sim 5-7$\%) variations in exergy and conversion efficiency impact the overall habitability of a planet? This is a challenging question but the asymptotic cases may be useful. For example, we can consider the hypothetical case of extreme super-habitable worlds (\citet{Heller2014}) in which all accessible chemical energy is utilized by life and phototrophic energy use is maximized (i.e. the biosphere cannot expand further from its equilibrium state). In this instance, two otherwise identical super-habitable worlds at these internal limits but hosted by a cooler and a hotter star respectively must exhibit different equilibrium biosphere sizes, due to the difference in photon conversion efficiency (Figures 1 \& 2). In this example we would expect the cooler star to host a smaller biosphere. That difference, in this specific case, is entirely a consequence of the 2nd law of thermodynamics, and is therefore `non-negotiable'.

At the other extreme, for planets at the very edge of habitability (either due to location in the CHZ or severe restrictions on niche environments or chemical availability) the difference in conversion efficiency between hot and cool stellar hosts could dictate the long-term survival of any phototrophic life at all. Of course, it could be argued that Darwinian selection may discover higher efficiency phototrophic chemical tools in extreme conditions, or with different stellar hosts with different spectra that induce different planetary photochemistry equilibria. But the selective forces acting on living systems will still be constrained by the fundamental limits on conversion efficiency quantified via exergy. 

In addition to these extremes, the greatest sensitivity of a biosphere (or components thereof) to $\sim 5-7$\% differences in conversion efficiency might occur at any ecological `tipping point'. For example, it has been suggested that the great end-Permian mass extinction on Earth some 250Myr ago would have exhibited strong latitude-dependency in terms of organisms that could respond to rapid marine hypoxia ($O_2$ depletion, \citet{Penn2018}). Equatorial species would be better adapted to warm waters and hypoxic conditions, whereas high-latitude species might not adjust as local marine conditions changed. But if there was an overall exergy advantage on an equivalent planet with a hotter parent star, that world might have a critical advantage in the face of dramatic geophysical fluctuations like those associated with the end-Permian, especially given the strongly non-linear nature of an event like this. Similarly, phototrophic organisms with lower local $T_s$ might also have an intrinsic advantage. In other words, hypothetically, cold-adapted high-latitude phototrophic organisms could benefit greatly from only slightly higher conversion efficiency and avoid mass die-off.

Furthermore, on the modern Earth, the depletion of rainforests - in particular those of the Amazonian basin - is predicted to have severe, and long-lasting impact on the overall climate state of the planet (e.g. see \citet{Lenton2011}). A variation in conversion efficiency at the level of 5-7\% between otherwise identical planets could translate into differences in surface area coverage by photoautotrophic life that result in highly non-linear, and significant differences in overall climate state between worlds.

Ultimately, both the ``quantity" (fluxes) and the ``quality" (thermodynamic efficiency) of available energy are important when analyzing possible biospheres. Notably, CHZ planets around low-mass (M-dwarf) stars not only receive less exergy with lower conversion efficiency, they receive comparatively lower fluxes of photosynthetically active radiation (see above) with respect to the Earth (e.g. \citet{Lehmer2018}, \citet{Lingam2019}). Thus it may be that M-dwarf exoplanets are doubly hampered in terms of sustaining complex biospheres based on photoautotrophy.

It is also worth noting that while this present work has focused on exergy in blackbody radiation (in part due to its relatively simple analytic form) there are other applications of Eqn 1. to the question of biospheres. For example, icy planets or moons with subsurface oceans could conceivably sustain life. Exergy and conversion efficiencies of biogeochemical pathways could be informative tools for examining different rocky-core/ocean configurations thermal conditions. Similarly, it would be interesting if a global exergy and efficiency budget could be derived for a place like Titan in our own solar system.

Future studies utilizing radiation exergy estimates for exoplanets would include investigating habitability joint-probabilities that combine the classic CHZ with exergy and efficiency functions. Also, further in the future, we might look for correlations between large populations of rudimentary spectral measurements of rocky planets (e.g. yielding low-precision temperature and atmospheric component constraints) and exergy properties. If the stacked data of planets with biospheres presents statistically significant differences from other population subsets (i.e. independent of the detailed variances of individual planets) this could be one way in which we convince ourselves of the presence of life on these worlds. If there are correlative trends towards, for example, higher conversion efficiency ($W^{BB}$) on candidate biosphere carriers, that would provide additional evidence supporting the positive detection of life. This would be especially important if those worlds are very different than the Earth (e.g. \citet{Seager2013}).

\section{\bf Acknowledgements}
The author thanks Frits Paerels for helpful discussions and comments and acknowledges support from the NASA Astrobiology Program through participation in the Nexus for Exoplanet System Science and NASA Grant NNX15AK95G. An anonymous referee is thanked for comments that have greatly improved the manuscript.


\begin{thebibliography}{}

\bibitem[Candau(2003)]{Candau2003} Candau, Y.\ 2003, Solar Energy, 75, 241
\bibitem[Chen(2005)]{chen2005} Chen, G. Q.\ 2005, Ecological Modeling, 184, 363
\bibitem[Delgado-Bonal(2017)]{Delgado2017} Delgado-Bonal, A.\ 2017, Nature Scientific Reports, 7, 1642
\bibitem[Dole(1964)]{Dole1964}Dole, S.~H.\ 1964, New York, Blaisdell Pub.~Co.~[1964] [1st ed.]., available online $https://www.rand.org/content/dam/rand/pubs/$
$commercial_books/2007/RAND_CB179-1.pdf$
\bibitem[Driscoll \& Barnes(2015)]{Driscoll2015}Driscoll, P.~E., \& Barnes, R.\ 2015, Astrobiology, 15, 739 
\bibitem[Hart(1979)]{Hart1979}Hart, M.~H.\ 1979, \icarus, 37, 351 
\bibitem[Heller \& Armstrong(2014)]{Heller2014} Heller, R., \& Armstrong, J.\ 2014, Astrobiology, 14, 50 
\bibitem[Jasechko et al.(2013)]{jasechko2013} Jasechko, S. et al.\ 2013, Nature, 496, 347 
\bibitem[Kaltenegger \& Sasselov(2011)]{Kaltenegger2011}, Kaltenegger, L., \& Sasselov, D.\ 2011, Astrophys. J. Lett., 736, L25
\bibitem[Kasting et al.(1993)]{Kasting1993} Kasting, J.~F., Whitmire, D.~P., \& Reynolds, R.~T.\ 1993, \icarus, 101, 108 
\bibitem[Keenan(1951)]{Keenan1951} Keenan, J.~H.\ 1951, British Journal of Applied Physics, 2, 183
\bibitem[Kiang et al.(2007a)]{Kiang2007a} Kiang, N.~Y., Siefert, J., Govindjee, \& Blankenship, R.~E.\ 2007, Astrobiology, 7, 222 
\bibitem[Kiang et al.(2007b)]{Kiang2007b} Kiang, N.~Y., Segura, A., Tinetti, G., et al.\ 2007, Astrobiology, 7, 252 
\bibitem[Kopparapu et al.(2013)]{Kopparapu2013} Kopparapu, R.~K., et al.\ 2013, \apj, 765, 131
\bibitem[Lehmer et al.(2018)]{Lehmer2018} Lehmer, O.~R., Catling, D.~C., Parenteau, M.~N., and Hoehler, T.~M.\ 2018, \apj, 859, 171
\bibitem[Lenton(2011)]{Lenton2011} Lenton, T.~M.\ 2011, Nat. Clim. Change, 1:201-209
\bibitem[Lingam and Loeb(2019)]{Lingam2019} Lingam, M., Loeb, A.\ 2019, arXiv:1901.01270
\bibitem[Loeb(2014)]{loeb2014} Loeb, A.\ 2014, International Journal of Astrobiology, 13, 337
\bibitem[Newman \& Sagan(1981)]{SaganNewman1981} Newman, W.~I., \& Sagan, C.\ 1981, \icarus, 46, 293 
\bibitem[N\"{u}rnberg et al.(2018)]{nurnberg2018} N\"{u}rnberg, D. J.\ 2018, Science, 360, 1210
\bibitem[Penn et al.(2018)]{Penn2018} Penn, J.~L., Deutsch, C., Payne, J.~L., Sperling, E.~A.\ 2018, Science, 362,  Issue 6419, eaat1327 
\bibitem[Petela(1964)]{Petela1964} Petela, R.\ 1964, Journal of Heat Transfer, 86, 187
\bibitem[Planck(1914)]{Planck1914} Planck M. The theory of heat radiation. P. Blakiston’s Son \& Co.,
Philadelphia, PA, 1914, http://gutenberg.org/ebooks/40030.
\bibitem[Ramirez(2018)]{Ramirez2018} Ramirez, R.~M.\ 2018, Geosciences, 8, 280 
\bibitem[Rant(1956)]{Rant1956} Rant, Z.\ 1956, Forschung Ing.-Wesens, 22, 36
\bibitem[Rosen (1954)]{Rosen1954} Rosen, P.\ 1954, Phys. Rev., 96, 555
\bibitem[Seager(2013)]{Seager2013} Seager, S.\ 2016, Science, 340, 577
\bibitem[Szargut et al.(1988)]{Szargut1988} Szargut, J., Morris, D., Steward, R.\ 1988, Exergy Analysis of Thermal, Chemical, and Metallurgical Processes. Hemisphere Publishing Corporation, New York.

\end{thebibliography}
\end{document}